# Decoding the shift-invariant data: applications for band-excitation scanning probe microscopy

Yongtao Liu,[1] Rama K. Vasudevan,[1] Kyle Kelley,[1] Dohyung Kim,[2] Yogesh Sharma,[3] Mahshid Ahmadi,[2] Sergei V. Kalinin,[1,a] and Maxim Ziatdinov[1,4,b]

[1] Center for Nanophase Materials Sciences, Oak Ridge National Laboratory, Oak Ridge, TN 37831, USA

[2] Joint Institute for Advanced Materials, Department of Materials Science and Engineering, University of Tennessee, Knoxville, TN 37996, USA

[3] Center for Integrated Nanotechnologies (CINT), Los Alamos National Laboratory, Los Alamos, NM 87545, USA

[4] Computational Sciences and Engineering Division, Oak Ridge National Laboratory, Oak Ridge, TN 37831, USA

A shift-invariant variational autoencoder (shift-VAE) is developed as an unsupervised method for the analysis of spectral data in the presence of shifts along the parameter axis, disentangling the physically-relevant shifts from other latent variables. Using synthetic data sets, we show that the shift-VAE latent variables closely match the ground truth parameters. The shift VAE is extended towards the analysis of band-excitation piezoresponse force microscopy (BE-PFM) data, disentangling the resonance frequency shifts from the peak shape parameters in a model-free unsupervised manner. The extensions of this approach towards denoising of data and model-free dimensionality reduction in imaging and spectroscopic data are further demonstrated. This approach is universal and can also be extended to analysis of X-ray diffraction, photoluminescence, Raman spectra, and other data sets.

[a] sergei2@ornl.gov
[b] ziatdinovma@ornl.gov

Following the invention of Atomic Force Microscopy (AFM) in 1986,[1] scanning probe microscopy (SPM) has emerged as an extremely powerful for probing and modifying nanoscale systems. Multiple variants of SPM methods were developed for probing electric,[2, 3] mechanical,[4, 5] electromechanical,[6-8] and magnetic phenomena,[9] as well as probing electronic[10, 11] and ionic transport.[12] Despite the broad gamut of measured signals and imaging conditions, the basic detection principles of these techniques remained almost invariant for two decades, and were based either on detection of static signal in e.g., contact AFM, amplitude detection via lock-in amplifier, or frequency detection via phase locked loop. In all these cases, the dynamic response of the cantilever is essentially reduced to one or several parameters linked to the functionality and visualized as spatially resolved maps.

This paradigm changed with the development of band excitation (BE) method,[13] which utilized parallel detection of response in multiple frequencies. BE has enabled quantitative measurements avoiding the frequency-dependent cross-talk in a broad variety of SPM methods including piezoresponse force microscopy (PFM),[14] Kelvin Probe Force Microscopy,[15, 16] and magnetic force microscopy.[17, 18] It has also enabled new SPM modalities including electrochemical strain microscopy[19] and electrochemical force microscopy.[20] However, the central element in BE, i.e., conversion from the measured amplitude/phase – frequency dependence to local dynamic characteristics has remained unchanged since its inception.[13] Namely, the response curve is fitted by a simple harmonic oscillator (SHO) model, and the derived response amplitude, resonance frequency, and quality factor are visualized as a function of spatial coordinates or control parameters such as voltage and time in complex spectroscopies.

Recently, deep neural networks were proposed as a way to improve the initial guesses to BE analyses.[21] However, this approach still postulates the SHO functional form of response, ignoring more complex mechanical and nonlinear responses. A Bayesian inversion approach was proposed to separate linear and non-linear responses.[22] However, this approach requires a prior set of models and is extremely computationally intensive. Hence, of interest is the development of unsupervised machine learning methods capable of analysis of BE data, and extendable to other similar data sets such as those emerging in X-Ray scattering, mass-spectrometry, or optical and Raman spectroscopies.

The common characteristic of these types of data sets is the presence of sharply localized peaks that can shift across the horizontal axis. For BE-SPM, this is due to variation of contact resonance frequency, for X-Ray – due to change in solid solution compositions and strain, etc. This behavior makes the applications of the linear multivariate methods such as principal component analysis (PCA) and related techniques[23] impractical, since decomposition gives rise to a large number of components. In BE, this was recognized since the first application of PCA in 2009.[24] This behavior persists for more complex analysis methods, including those based on the manifold learning, conventional and variational autoencoders (VAE).[25]

Here, we propose and implement a novel manifold learning method based on a shift-invariant VAE (Figure 1a). This approach allows naturally accounting for the shift of peaks along the stimulus axis. We show that under certain conditions, the latent variables derived from the unsupervised learning are linear functions of the ground truth peak shift and other ground truth parameters of the peaks, corresponding to the full unsupervised disentanglement of physically-relevant variables. Here, the application for band-excitation PFM is developed. However, extensions for other methods such as X-Ray scattering, photoluminescence, and Raman spectra are straightforward.

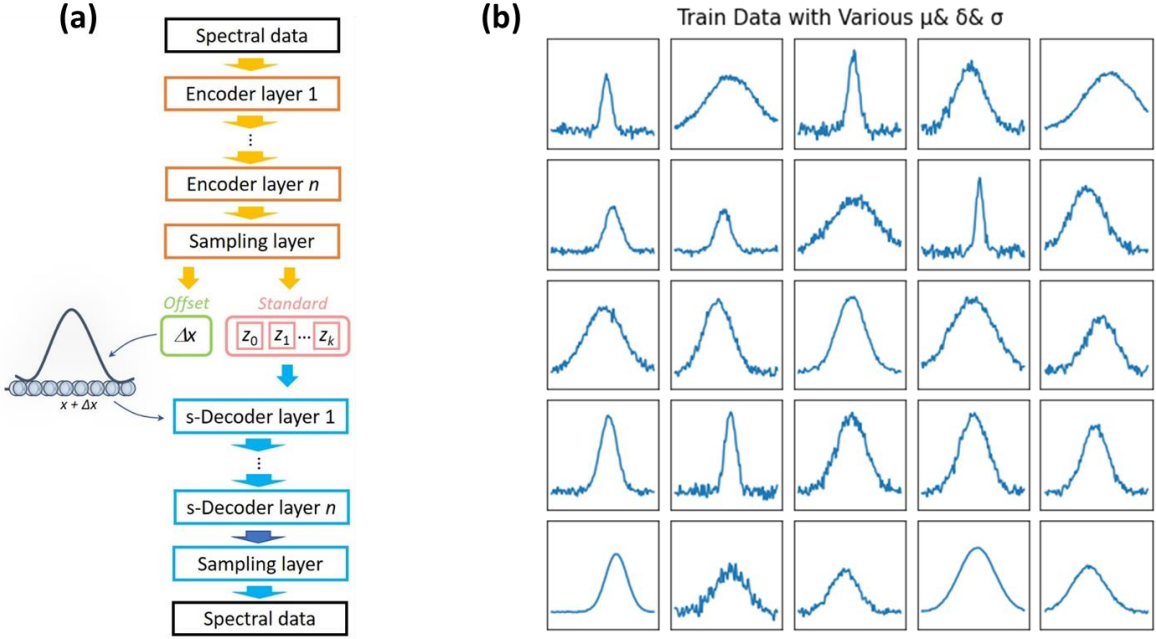

**Figure 1. (a),** Schematic of shift-VAE. The encoder maps the input spectral data into the offset latent variable ($\Delta x$) and conventional VAE latent variables ($z_0$, …, $z_k$). The former is used to shift the coordinate grid which is then concatenated with the remaining latent variables and passed to the decoder (which is now a function of coordinates). We then score the observed data against the Gaussian likelihood parametrized by the decoder output. Both the encoder and decoder are multi-layer perceptrons with *tanh( )* non-linear activations. **(b),** a representative synthetic 1D spectra data set, this figure only randomly shows 25 spectra in the data set, however, this data set is a collection of 5000 1D Gaussian curves with shift, $\mu \in [-3, 3]$, intensity, $\sigma \in [0.5, 1]$, and width $\delta \in [0.5, 5]$.

To illustrate the origins of this problem, we illustrate analysis on synthetic data set. Here, the data set is formed as a collection of $N = 5000$ Gaussian curves defined on $x \in [-10, 10]$ by function:

$$y = \sigma \times e^{\frac{-(x-\mu)^2}{2 \times \delta^2}} + \sigma_{noise}$$

with the shifts uniformly distributed on $\mu \in [\mu_{min}, \mu_{max}]$, amplitudes on $\sigma \in [\sigma_{min}, \sigma_{max}]$, and width on $\delta \in [\delta_{min}, \delta_{max}]$. In addition, the white noise of amplitude $\sigma_0$, or amplitude uniformly distributed on $\sigma_{noise} \in [0, \sigma_{max-noise}]$ interval, can be added. The data can be optionally normalized. Figure 1b shows a representative synthetic data set with $\mu \in [-3, 3]$, $\sigma \in [0.5, 1]$, and $\delta \in [0.5, 5]$. More example of synthetic data sets are shown in Supplementary Materials (Figure

S1) and the provided notebook allows adjusting the synthetic data set parameters and subsequent analytics.

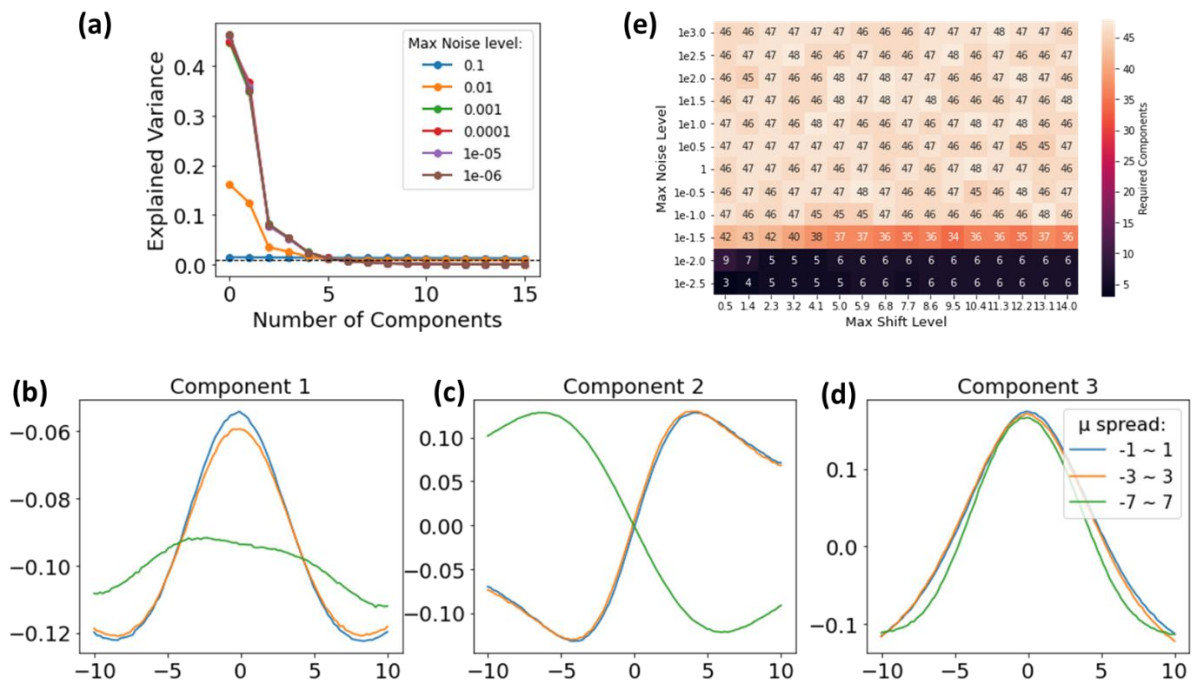

**Figure 2. Principal component analysis (PCA) of synthetic data set.** **(a),** scree plots of the synthetic data sets with different noise levels. **(b-d),** first three PCA components of the data sets with different width spread. **(e),** heat map showing the number of PCA components required to represent 99% of the data when noise and peak width vary, the required number of components is labeled in the map.

The PCA scree plots of the synthetic data is shown in Figure 2a for different noise levels $\sigma_{noise} \in [0, \sigma_{max-noise}]$, $\sigma_{max-noise} = 10^{-1}, 10^{-2}, 10^{-3}, 10^{-4}, 10^{-5}$, or $10^{-6}$. Here, the curve illustrates characteristic shape with rapid decay of the values followed by the saturation in the white noise tail. The inflection point separating the two regimes defines a number of significant components required to fully represent the data. The characteristic PCA components are shown in Figure 2b-d and illustrate the average and factor of variability in the degree of importance. Note that these components, generally speaking, do not have a well-defined physical sense. Finally, a heat map in Figure 2e shows the number of components required to represent the data as a function of the maximum noise level and maximum shift $\mu_0$. Here, the number of components is estimated as components needed to represent 99% of information in data set. Note that for large distributions or low noise levels the number of components can be very significant. These behaviors were observed experimentally, with often highly non-trivial distribution of spatial information between components. Shown in Figure S2 is an analysis of the correlation between PCA components and Gaussian curves' properties (i.e., $\mu, \sigma, \delta$), where none of the components shows a correlation with curves properties, clearly illustrating limitations of the PCA or other linear analysis methods for such data sets.

To alleviate this problem, we introduce a shift-VAE technique. The key idea behind a regular VAE is that complicated real-world observations can be explained by a small number of disentangled latent variables capturing the ground truth factors of variation. The VAE consists of a decoder (generative model) that reconstructs observations from a latent code and an encoder (inference model) that approximates the true posterior probability via the amortized variational inference. Here, it is important to note that Locatello *et al.* demonstrated[26] (theoretically) that unsupervised learning of disentangled latent representations is fundamentally impossible without inductive biases. The latter typically involves modification of the loss function and/or of the architecture of the encoder and decoder neural networks.[26, 27] Here, we argue that for applications of VAE (and other deep generative models) in domain sciences, the necessary inductive bias(es) can come from prior domain knowledge such as understanding the role of measurement (instrumental factors) and the information available from theory about fundamental length scales and symmetries present in a system. This concept is illustrated by the shift-VAE technique for analyzing 1D spectral data in the presence of arbitrary shifts in peak position.

In shift-VAE (Figure 1a), we designate one of the latent variables to absorb the information about the relative position of spectral features (we refer to it as "offset" latent variable), whereas the rest of the latent variables capture other (than position) factors of variation. Specifically, we start by creating a 1D *x*-coordinate grid whose length is equal to the number of points in the spectra. Our encoder maps the input spectra into the offset latent variable and several (usually two) standard VAE latent variables. We assume that shifts in the position are normally distributed and sample our offset latent vector from a Gaussian distribution, although the usage of other distributions is also possible. The sampled values are used to shift the 1D coordinate grid by *Δx*. The shifted grid is then concatenated with the standard latent vector *z* and passed to the VAE's generator network expressed as a function of spectral coordinates to enforce consistency in geometric features between the shifted spectra. Practically, we multiply the offset latent vector by a coefficient *k* ($0 < k \leq 1$) whose value reflects our prior belief about a degree of "disorder" in the system. The loss (negative evidence lower bound, ELBO) is computed according to

$$\mathcal{L}(y) = RE + \beta_1(t)D_{KL}\big(q(z|y)\|p(z)\big) + \beta_2(t)D_{KL}(q(\Delta x|y)\|p(\Delta x)),$$

where *RE* is a reconstruction error, $D_{KL}$ is a Kullback-Leibler divergence term, and *β* are (optional) "time"-dependent regularization coefficients (here "time" is expressed through a training iteration number).

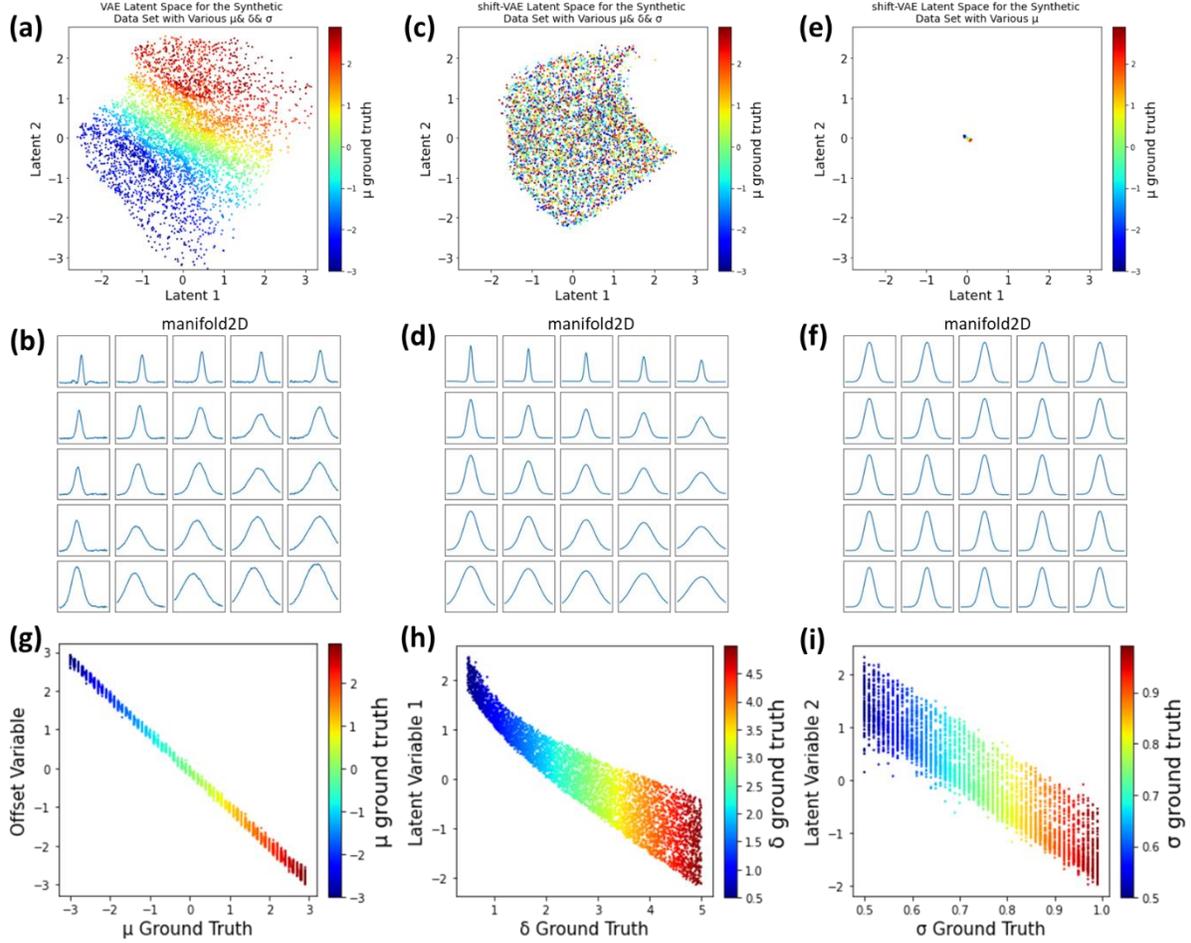

**Figure 3.** The VAE and shift-VAE analysis results presented as scatter plots of the encoded data points in the latent space and a 2D latent manifold projected to the input (spectral) space. **(a), (c),** latent variables distribution for VAE and shift-VAE encoding of the data from Figure 1b; in this dataset the curve parameters vary as following: $\mu \in [-3, 3]$, $\sigma \in [0.5, 1]$, and $\delta \in [0.5, 5]$; **(b), (d),** corresponding learned latent manifolds. **(e), (f),** shift-VAE analysis of the data set with only one factor of variability, shift $\mu \in [-3, 3]$, whereas both width and intensity are fixed: $\sigma = 1$ and $\delta = 2$ (see Figure S1a); given that shift-VAE separates the relative peak shift into a specific offset variable, both conventional latent variables are collapsed. **(g-h),** the relationships between the ground truth values and latent variables derived from the shift-VAE analysis in **(c-d)**. It is seen that there is a linear correlation between ground truth $\mu$ and the offset variable **(g),** while the information regarding peak width $\delta$ and intensity $\sigma$ is disentangled into latent variable 1 **(h)** and latent variable 2 **(i),** respectively. More information of the relationship between ground truth values and latent variables of VAE and shift-VAE analysis can be found in Supplementary Information Figure S3 and S4.

Shown in Figure 3 are the results of the VAE and shift-VAE analysis. Note that, compared to shift-VAE, VAE technique does not have an offset latent variable, so the relative position of spectra is captured in conventional latent variables. Here, the curves in the original data set (Figure 1b) are encoded into two latent variables, and resultant distribution is plotted on the 2D plane. Shown in Figure 3a is the latent space distribution for the VAE encoding. The distribution is reminiscent of the ground-truth distributions of widths and positions. The color scale corresponds to the ground truth value of peak shift, $\mu$. Notable is that the labels corresponding to the position are changing (mostly) from top to bottom of the image. The labels corresponding to the width are changing (mostly) from left to right (not shown). We further reconstruct the curves (Figure 3b) from the uniform square grid of points in the latent space and observe that indeed the width changes from top to bottom and position changes from left to right. Hence, our VAE has (mostly) disentangled the representations of the data.

We also explored the correlation between the VAE's latent variables and the ground truth and found that there is indeed a certain degree of disentanglement between the two. In other words, the latent representations of the data disentangled by VAE show the factors of variability in the original synthetic data set. However, while there is a relationship between the ground truth values and latent variables, they are not equal and have large deviations (Figure S4). Furthermore, the offsets are encoded in arbitrary units, which is not of practical use. Most importantly, the VAE disentanglement works well for narrow distributions, for which the peaks are fully confined within the data interval. For broad distributions, the shape of the latent manifold changes, and the correlation between the ground truth and the disentangled representation breaks down. In other words, the VAE behavior becomes controlled by the cut-off at the edges of the interval, which can be seen from the Supplementary Video 1 of the VAE latent space evolution when $\delta$ of training data set gradually changes from [0.5, 1] to [0.5, 50].

Next, we analyzed the shift-VAE results. Shown in Figure 3c-d is the shift-VAE analysis of the same data set (see Figure 1b). In this case, the relative shift is separated as a special offset latent variable, and the remaining variability is encoded as two conventional latent variables. In this case the offset variable solely represents peak shift ($\mu$) and two conventional latent variables encode the other parameters. Indeed, shown in Figure 3c is the distribution of the two latent variables encoded by shift-VAE, with the color corresponding to the ground truth shift. The latter is randomly distributed, clearly indicating that the shift variability was separated from the other latent variables. This is also illustrated for the learned latent manifolds projected to the input (spectral) space (Figure 3d), where we only observe peak width changes, and the peak position is the same for all the curves. In Figure 3g-i we explore the relationships between the encoded latent variables of the shift-VAE and the ground truth values. We observe that the offset variable shows a clear linear relationship with the ground truth $\mu$ (Figure 3g). Here, of particular importance is that the absolute values of the offset variable and ground truth $\mu$ are equal. Moreover, the offset variable is independent of ground truth width ($\delta$) and intensity ($\sigma$), as shown in Figure S4c and e, respectively. This indicates that the shift-VAE disentanglement works very well for encoding peak shift. In addition, variabilities of width ($\delta$) and intensity ($\sigma$) are also mostly encoded into separated latent variables by shift-VAE, as shown in Figure 3h and 3i, respectively. Most importantly, shift-VAE also performs considerably well for broad peaks, where even if partial peaks are cut off at edges of the interval, as shown in Figure S5.

Note that for data sets with fewer factors of variability, the latent space will be partially or even completely collapsed. For example, shown in Figure 3e-f are latent space distribution and the learned latent manifolds of a shift-VAE analysis on the data set in Figure S1a with only one factor of variability, shift µ, for which both latent variables are collapsed. This is because the only existing variability factor (µ) is captured by the offset variable and hence conventional latent variables are variability-free in this case. In contrast, the latent space of the VAE analysis (shown in Figure S6a) on this data set is only partially collapsed, with the second latent variable being related to the ground truth shift. Classically in VAEs, the collapse of latent space is perceived as a problem calling for the adjustment of the "loss" function. However, in our case, we know that the ground truth data set has only limited factor of variability. Hence, the dimensionality of our latent space hints at the true physical dimensionality of the data. This is further confirmed by analyses on the data set with two factors of variability, as illustrated in detail in Supplementary Figure S6.

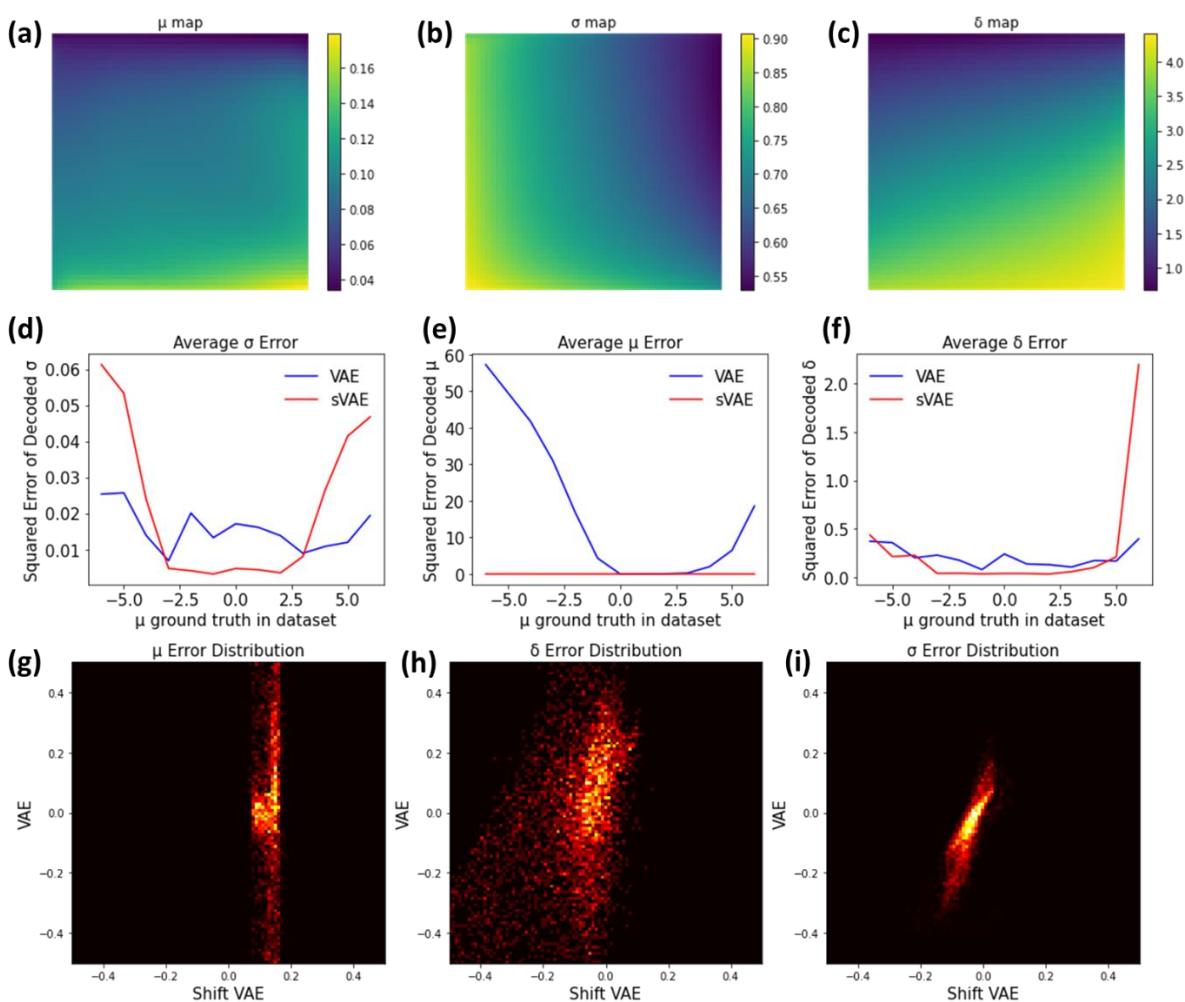

**Figure 4**. **(a-c),** shift-VAE maps of peak parameters (σ, µ, δ) derived from the learned manifold data projected to the spectral space. **(d-f),** VAE and shift-VAE decoded peak parameter errors as a function of the ground truth shift of the analyzing datasets. **(g-i),** comparing the performance of VAE and shift-VAE by plotting the errors of decoded peaks as 2D hist map.

To obtain further insight into this behavior, we explore the structure of the latent space of VAE and shift-VAE in terms of function behavior. Here we fit the curves reconstructed from the latent manifold by the known ground truth functional form (*i.e.,* Gaussian) and show the maps of fit parameters σ, μ, δ in Figure 4a-c (shift-VAE) and Figure S7a (VAE). In Figure 4a-c, we observe that μ maps of shift-VAE are almost uniform and all values are very close to zero. This is because that the shift-VAE separates the peak shift into the offset variable and hence the reconstructed curves from shift-VAE centered at $\mu_0$, which is the center of μ for the train data set (for this training data set with $\mu \in [-3, 3]$, the $\mu_0$ is zero). In this case, the conventional latent variables only represent other variability factors, such as intensity (σ) and width (δ). Therefore, we observe gradual changes from left to right in σ map (Figure 4b) and from top to bottom in δ map (Figure 4c), indicating intensity is encoded in the first latent variable (horizontal) and width is encoded in the second latent variable (vertical). The additional analyses of VAE and shift-VAE for data sets with different ground truth parameters are given in Supplementary Materials Figure S7 and Figure S8.

To quantify the behavior of VAE and shift-VAE, we applied the trained VAE and shift-VAE models to analyze data sets with known ground truth values. These data sets can be generated by setting desired ground truth values, which can be done through the provided Jupyter Notebook. For example, in this analysis, we used 13 data sets with specified ground truth μ. For each data set, the ground truth μ is fixed at $\mu_s$ ($\mu_s \in [-6, 6]$ and $\mu_s$ is an integer). We first encode the raw data sets with trained VAE and shift-VAE models, then reconstruct the spectra using encoded latent vectors. Afterward, we fit the reconstructed spectra by Gaussian function and compare the fit parameters with ground truth values. The difference between fit parameters and ground truth values is taken as the error. Shown in Figure 4d-f are the average errors of VAE and shift-VAE analyses on each data set as a function of the ground truth μ of the data set. Clearly, shift-VAE behaves well in encoding peak shift, where the shift error of shift-VAE is always close to zero, as shown in Figure 4e. In contrast, VAE only behaves well for the data set with ground truth shift μ near zero. The behavior of VAE and shift-VAE is also compared by plotting the 2D hist maps of errors, as shown in Figure 4g-i. It indicates that for all errors (μ, δ, and σ) the spread is larger in the VAE direction than the shift-VAE direction, suggesting the better performance of shift-VAE.

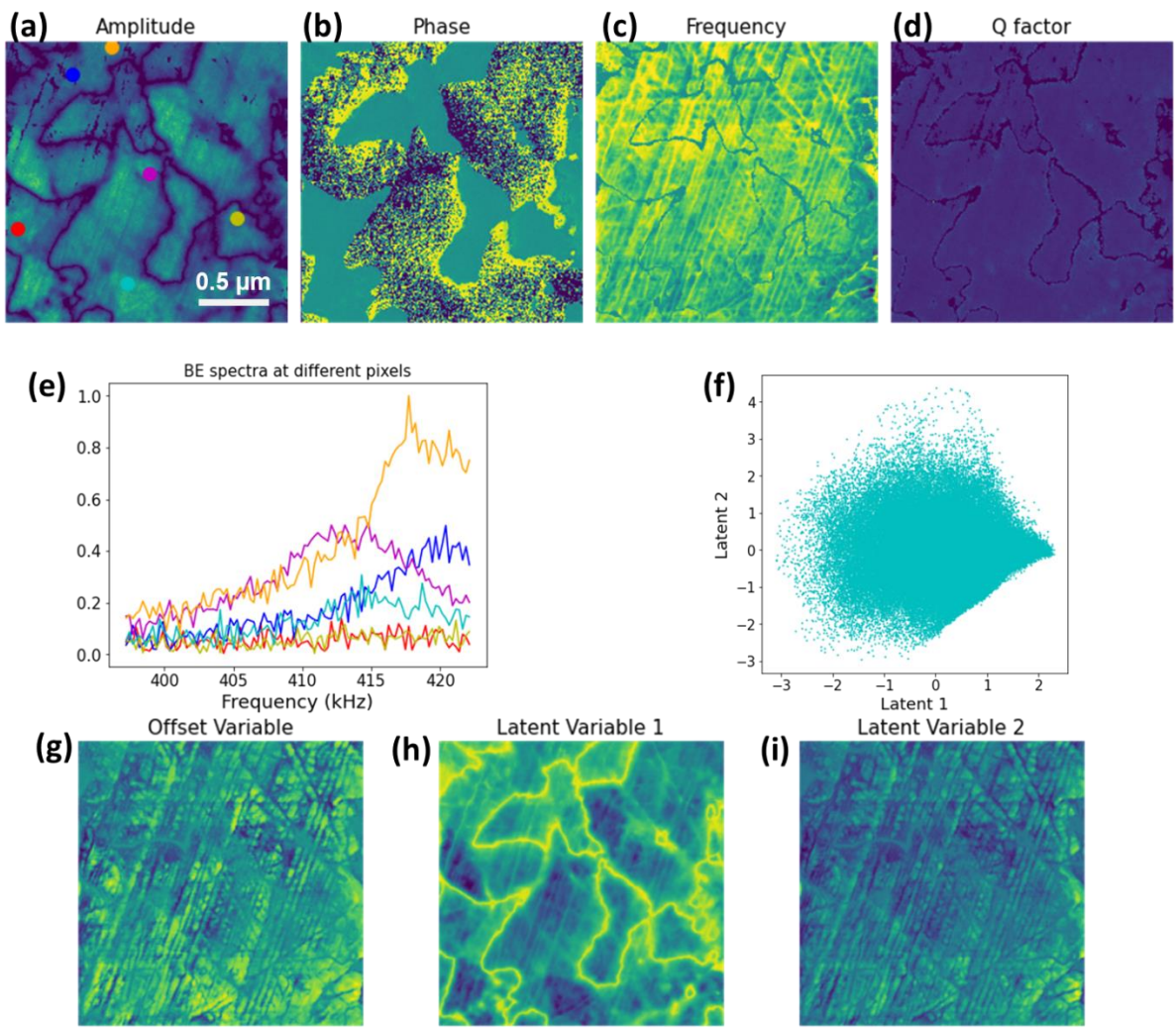

**Figure 5**. **BE-PFM data of BFO and shift-VAE analysis latent space. a-d,** BE-PFM results of a BFO sample show ferroelectric domain structure. **e,** BE spectra from several random locations. **f,** latent space of shift-VAE analysis on this BE-PFM data. **g-i,** maps of the offset variable and latent variables of shift-VAE analysis.

We further proceed to extend this analysis to experimental data sets. Shown in Figure 5a-d is band excitation (BE) PFM data from bismuth ferrite (BFO) bulk ceramics, for which signals of 180º ferroelectric domains mainly show up in amplitude and phase images. In addition to ferroelectric domain contrast, frequency images also show some scratch stripes associated with the resonance frequency shift likely due to crosstalk. In Figure 5e, we show several BE spectra from random locations (as marked in Figure 5a), which indicate the variability of peak intensity, width, and position among these spectra.

The VAE and shift-VAE analyses are then performed on this BE-PFM data to disentangle information contained in BE spectra. Shown in Figure S9 are the VAE results for latent dimensionality of two, for which the first latent variable well represents the ferroelectric domain structure, and the second latent variable represents the crosstalk scratch lines. Latent space starts to collapse when increasing the number of latent variables (i.e., latent space dimensionality), as shown in Figure S10-S11 for latent dimensionality equals of three and four. Nonetheless, the ferroelectric domain structure can always be disentangled into a non-collapsed latent variable. The results of shift-VAE analysis for latent dimensionality of two are shown in Figure 5f and the maps of the offset and latent variables are depicted in Figure 5g-i. In this case, the offset variable map (Figure 5g) mainly shows the scratch line structure. Similar to VAE, the latent space of shift-VAE starts to collapse with an increase in latent dimensionality, as shown in Figure S12-S13. This behavior is likely associated with the physical dimensionality of BE-PFM spectra, which ideally only contains information about resonance frequency (peak shift) and polarization amplitude (peak intensity). However, in a more realistic scenario, we can expect a slightly larger physical dimensionality of BE spectra due to measurement artifacts. Of importance is that for shift-VAE the offset variable always represents the crosstalk scratch lines. This matches our understanding of the physical mechanism of BE-PFM, for which crosstalk can induce resonance frequency changes (peak shift).

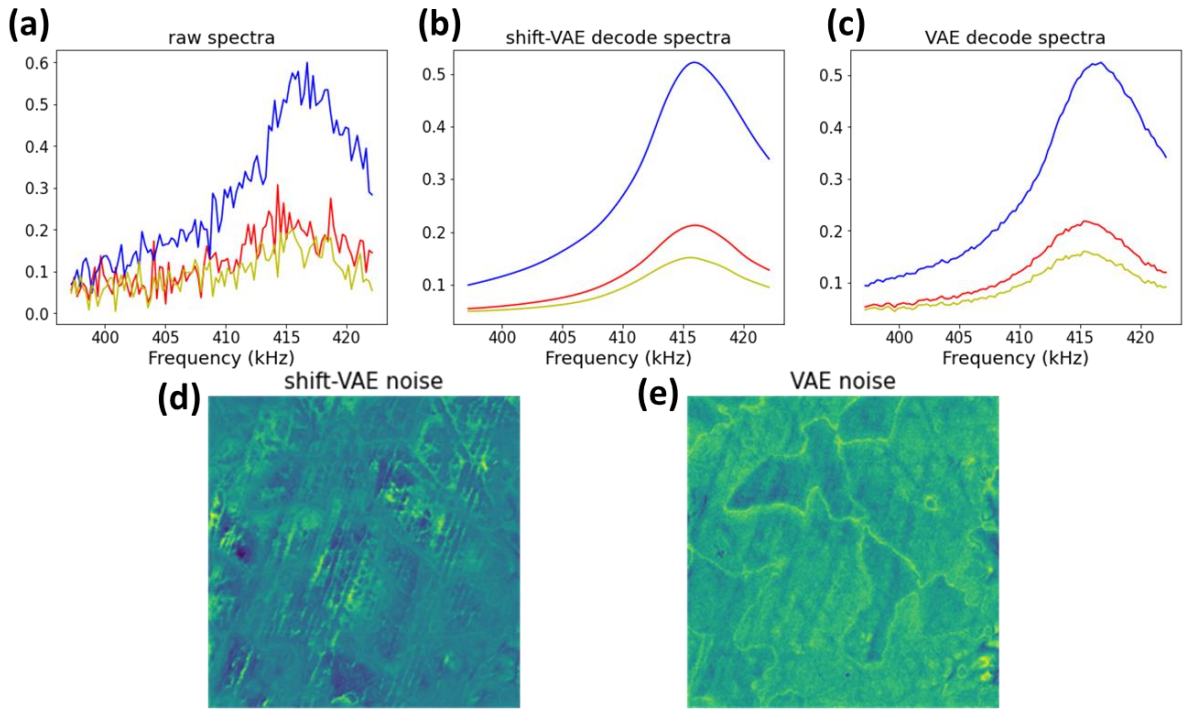

**Figure 6**. **shift-VAE and VAE reconstruction of BE spectra.** **(a), (b), (c),** spectra at three random locations, **(a),** raw spectra; **(b),** shift-VAE reconstructed spectra; **(c),** VAE reconstructed spectra. These spectra indicate the denoising capability of the shift-VAE and VAE techniques. **(d), (e),** calculated noise map by subtracting shift-VAE / VAE reconstructed spectra and raw spectra, which indicate whether shift-VAE / VAE analyses remove real information of the data. It is clear that VAE analysis partially removes the information regarding domain structure.

The remarkable aspect of VAE and shift-VAE analysis provides potential for systematic imputation on missing and high-noise data. For example, using a traditional SHO fit for data with the high noise leads to large uncertainty in resonance frequency position, necessitating the development of *ad hoc* criteria to mask the images based on amplitude or fit quality. Comparatively, VAE and shift-VAE approach allow to denoise the low-intensity curves using the information contained in high-intensity ones. This behavior is shown in Figure 6a-c of raw spectra, shift-VAE and VAE reconstructed spectra at three random locations. Clearly, both shift-VAE and VAE denoise the spectra, but shift-VAE shows a better performance. Even if for the spectra with low peak intensity (e.g., the yellow spectra), shift-VAE and VAE can reconstruct the spectra and reduce noise. In a denoising process, an important aspect we need to consider is whether any real information is removed during it. To evaluate this, we calculate the noise by subtracting reconstructed spectra and raw spectra, then show the noise as maps. Figure 6d and 6e show the noise maps calculated based on shift-VAE and VAE reconstructed spectra, respectively. Note that the VAE noise map shows ferroelectric domain structure, suggesting the denoising process of VAE removes a portion of real response. However, this does not occur in shift-VAE, again indicating the better performance of shift-VAE.

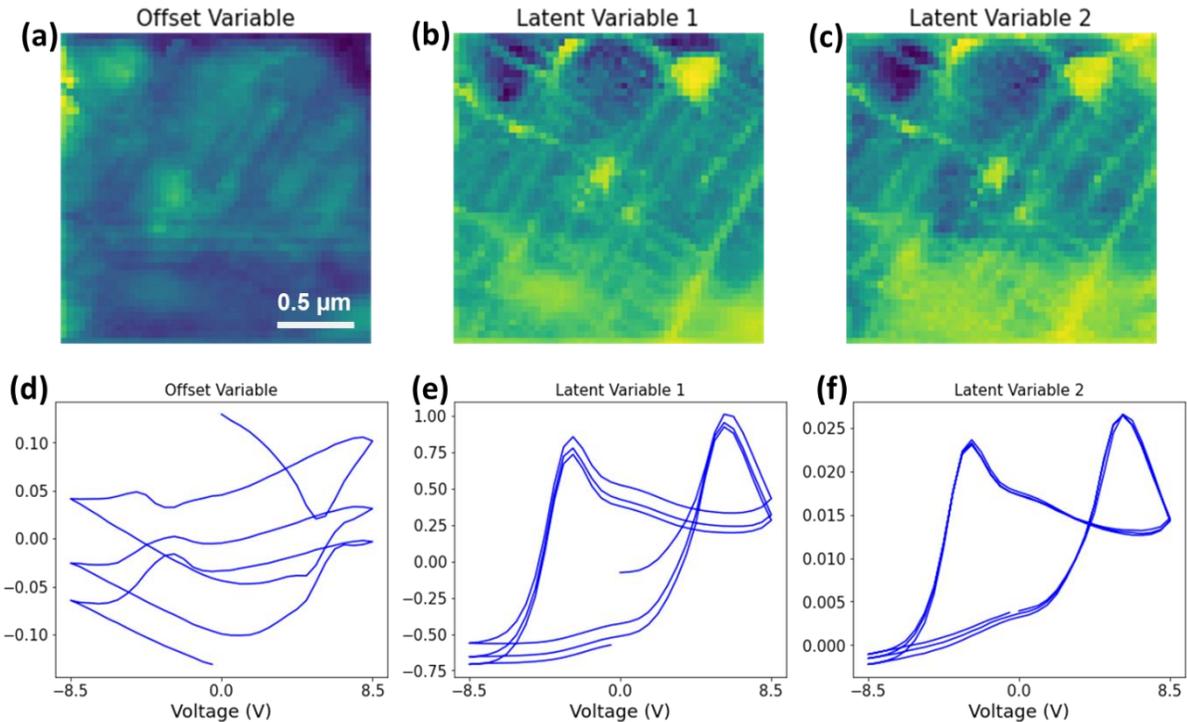

**Figure 7. shift-VAE results of spectroscopic BE-PFM data.** **(a), (b), (c),** maps of the offset variable and latent variables. **(d), (e), (f),** behavior of the offset variable and latent variables as a function of applied bias.

Finally, we extend this approach to analysis of the spectroscopic imaging in PFM, and by extension, more complex spectroscopies. As a model system, a lead titanate (PTO) sample is used to collect spectroscopic BE-PFM result. A bipolar triangular bias waveform is applied to switch the sample; basic PFM maps and switch loops are shown in Figure S14 and S15. We note that this PFM data was used in our earlier work;[28] here we just use this data as a model system for shift-VAE analysis. When analyzing this data by shift-VAE, the shift-VAE reconstructed spectra match well with the corresponding raw spectra (Figure S16), indicating the good performance of shift-VAE analysis on this data. Figure 7a-c shows the shift-VAE offset variable and conventional latent variables as 2D maps. We observe that domain structure is mainly captured by conventional VAE latent variables and the offset variable map only shows overall background information. Since this is spectroscopic data with switching information, further analysis is performed to look at the offset variable and latent variables as a function of the applied bias. In this analysis, the offset variable and latent variables are respectively averaged over the map at each bias step. Figure 7d-f shows the loops as a function of the applied bias. We observe that the offset variable shows a continuous drift over the switching process (Figure 7d). This drift is likely related to the resonance frequency shift induced by the material change or surface charge accumulation during the application of bias. Then, latent variables show inverse butterfly loops, which are similar to the amplitude butterfly loop for switching of ferroelectric materials. This analysis indicates that shift-VAE disentangles the information related to material failure or electrostatic response from the information regarding the ferroelectric response, further strengthening the powerfulness of shift-VAE in learning spectra data.

To summarize, we introduced a shift-invariant variational autoencoder (shift-VAE) for analyzing 1D spectra data in a model-free unsupervised manner, which enables naturally accounting for the properties of spectral data. Using synthetic Gaussian peak data sets, we show that the shift-VAE latent variables derived from the unsupervised learning are linear functions of the ground truth parameters, disentangling physically relevant variables. The application of shift-VAE is illustrated for band-excitation piezoresponse force microscopy (BE-PFM) data. From two BE-PFM data sets, we consistently observed that shift-VAE successfully disentangles ferroelectric polarization and crosstalk-induced resonance frequency peak shift from BE curves. In addition, shift-VAE also shows the strength in denoising raw curves without sacrificing real response, allowing the recognition of real response (even if) for low signal-to-noise ratio data. Overall, shift-VAE proves to be a powerful technique for learning spectra, which should have very broad applications in many fields.

**Materials and Methods**

The shift-VAE is realized using a home-built pyroVED package (https://github.com/ziatdinovmax/pyroVED). The details of shift-VAE and VAE analysis are available from Jupyter notebook at https://git.io/JOgFB.

$BiFeO_3$ (BFO) ceramic was synthesized by a conventional solid-state reaction method using $Bi_2O_3$ and $Fe_2O_3$ power precursor (Alfa Aesar >99.99%). The powder precursor is weighed in stoichiometric amounts. The powder mixtures were calcined at 760 °C for 1.5 h in air. After calcination, the powders were pressed using a cold isostatic method and then sintered at 780 °C for 1 h in air to form polycrystalline ceramics. The surface of the synthesized BFO bulk ceramic was polished for BE-PFM measurements. Pt-coated tips (Budget Sensors, ElectriMulti75-G; nominal resonance frequency, 75 kHz, 3N/m) were used to apply voltages to the AFM tip. The used drive amplitude was 2.5 V (AC bias) for BE-PFM excitation. The frequency range for BE-PFM was centered 410 kHz. The measurements were taken over a grid of 256*256 pixels on the sample surface.

The PTO film was grown by chemical vapor deposition on a $SrRuO_3$ bottom electrode on a $KTaO_3$ substrate.[29] The BE-PFM imaging and spectroscopy were performed on a commercial microscope (Cypher, Asylum/Oxford Instruments) at room temperature with a Pt/Ir-coated tip (1 N/m). In-house developed LabVIEW code was used to acquire the band-excitation piezoforce spectroscopy data using National Instruments hardware. DC voltage ramping from −12.0 to + 12.0 V was applied to the tip to measure the piezoresponse using band-excitation approach with a 1 V AC signal.[28]

**Conflict of Interest**

The authors declare no conflict of interest.

**Authors Contribution**

S.V.K. conceived the project and M.Z. realized (shift-) VAE in Pyro probabilistic programming language. Y.L. performed analyses. K.K., D.K., and R.V. performed BE-PFM measurements. Y.S. synthesized BFO samples. S.V.K., M.Z., and Y.L. wrote the manuscript. All authors contributed to discussions and the final manuscript.

**Acknowledgements**

This effort (ML and PFM) is based upon work supported by the U.S. Department of Energy, Office of Science, Office of Basic Energy Sciences Energy Frontier Research Centers program under Award Number DE-SC0021118 (Y.L., S.V.K.), and the Oak Ridge National Laboratory's Center for Nanophase Materials Sciences (CNMS), a U.S. Department of Energy, Office of Science User Facility (M.Z.). D. K. and M. A. acknowledge support from CNMS user facility, project number CNMS2019-272. Y. S. acknowledges the support from the G. T. Seaborg Fellowship (project number 20210527CR) and the Center for Integrated Nanotechnologies, an Office of Science User Facility operated for the U.S. Department of Energy Office of Science at Los Alamos National Laboratory. The authors are thankful to Prof. Hiroshi Funakubo (Tokyo Institute of Technology) for providing PTO samples.